\documentclass{elsart} 
\usepackage{epsfig} 
\usepackage{amstext}
\begin{document}
\bibliographystyle{elsevier}
\begin{frontmatter}
  \title{The Longitudinal Polarimeter at HERA}
  \author[Freiburg]{M.~Beckmann}, 
  \author[UM]{A.~Borissov}, 
  \author[Freiburg]{S.~Brauksiepe}, 
  \author[Freiburg]{F.~Burkart}, 
  \author[Freiburg]{H.~Fischer}, 
  \author[Freiburg]{J.~Franz}, 
  \author[Freiburg]{F.H.~Heinsius},
  \author[Freiburg]{K.~K\"onigsmann}, 
  \author[UM]{W.~Lorenzon\thanksref{address}},  
  \author[Freiburg]{F.M.~Menden}, 
  \author[UM]{A.~Most},
  \author[UM]{S.~Rudnitsky},
  \author[Freiburg]{C.~Schill},
  \author[Freiburg]{J.~Seibert},  
  \author[Freiburg]{A.~Simon} 

  \address[Freiburg]{Fakult\"at f\"ur Physik, Universit\"at Freiburg, 79104 
               Freiburg, Germany}

  \address[UM]{Randall Laboratory of Physics, University of Michigan, Ann 
               Arbor, Michigan 48109-1120}
  
  \thanks[address]{Corresponding author; email: lorenzon@umich.edu}

\begin{abstract}
The design, construction and operation of a Compton back-scattering laser
polarimeter at the HERA storage ring at DESY are described. The device
measures the longitudinal polarization of the electron beam between the spin
rotators at the HERMES experiment with a fractional systematic uncertainty of
1.6\,\%. A measurement of the beam polarization to an absolute statistical
precision of 0.01  requires typically one minute when the device is operated
in the multi-photon mode. The polarimeter also measures the polarization of
each individual electron bunch to an absolute statistical precision of 0.06 in
approximately five minutes. It was found that colliding and non-colliding
bunches can have substantially different polarizations. This information is
important to the collider experiments H1 and ZEUS for their future
longitudinally polarized electron program because those experiments use the
colliding bunches only.
\end{abstract}


\end{frontmatter}

\section{Introduction}
\label{section:intro}

In high-energy storage rings, electron (positron) beams can become
transversely polarized through the emission of synchrotron
radiation~\cite{ST64}. This process involves a small asymmetry in the
spin-flip amplitudes, which enhances the population of the spin state
antiparallel (parallel) to the magnetic bending field. The polarization
develops in time according to
\begin{equation}
P(t) = P_\infty\left(1-e^{-t/\tau}\right),
\label{eq Pt}
\end{equation}
where the asymptotic polarization $P_\infty$ and the time constant $\tau$ are
characteristics of the ring conditions. In the absence of depolarizing
effects, the maximum polarization theoretically achievable is
$P_{\rm{th}}=0.924$, and the rise-time constant, which depends on the bending
radius of the storage ring and the beam energy, is $\tau_{\rm{th}}=37$~min for
the HERA storage ring operated at an energy $E_e = 27.5$~GeV.

Depolarizing effects can however substantially reduce the maximum achievable
polarization. These intricate effects cannot generally be precisely
controlled, making it necessary to continuously measure the beam polarization.
The depolarizing effects also affect the actual rise-time, which scales with
$P_\infty$ according to 
\begin{equation}
\tau=P_\infty \left(\frac{\tau_{\rm{th}}}{P_{\rm{th}}}\right).
  \label{eq:Pmax}
\end{equation}
Thus for a typical beam polarization of 0.55, the rise-time is about 22~min. 

This article describes a polarization monitor at the HERA electron ring at
DESY, which is based on Compton scattering of circularly polarized photons
from an intense pulsed laser beam. This method for measuring the polarization
of stored electron beams\footnote{electron beams refer to both, electron and
positron beams for the remainder of this article} was suggested more than 30
years ago~\cite{BK69}, and has been employed at many laboratories~\cite{Bar96}
to measure the transverse polarization. Compton scattering has also been
employed at linear accelerators~\cite{linacs} to measure the longitudinal
polarization. In recent years, the NIKHEF~\cite{Nik98} and
MIT-Bates~\cite{Bates00} laboratories have developed Compton polarimeters to
monitor the longitudinal beam polarization in their storage rings. At DESY, a
Compton polarimeter~\cite{Bar93} had been constructed in 1992 to measure the
transverse polarization of the electron beam in the HERA West section. This
Transverse Polarimeter measured the electron beam polarization with an initial
fractional systematic uncertainty of 9\,\%, which has subsequently been
improved to 3.4\,\%~\cite{TPOL-3.4}. 

The Longitudinal Compton Polarimeter was added to obtain an independent and
more precise measurement of the beam polarization at HERA, with very different
systematic uncertainties and the capability to measure individual bunch
polarizations. It was commissioned during fall 1996, and provides a
measurement of the longitudinal beam polarization in the East section of  HERA
between the spin rotators~\cite{BS86} at the HERMES
experiment~\cite{HERMES-spectrometer}. 

\section{Polarized Compton Scattering}

\begin{figure}[htbp]
\centerline{\epsfig{figure=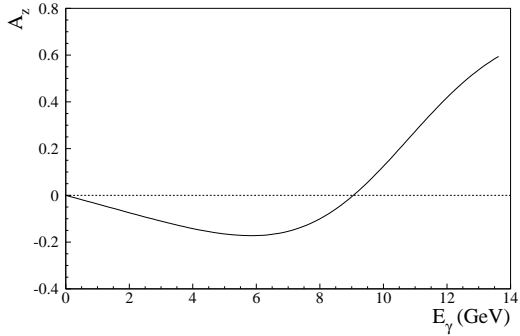,width=0.5\textwidth}}
\vspace*{1mm}
\caption{The longitudinal asymmetry function $A_z$ versus the energy
$E_\gamma$ of the back-scattered Compton photons for the case of a 2.33~eV
photon incident on a 27.5~GeV electron.}
  \label{fig:analyzing-power.eps}
\end{figure}

The cross section for Compton scattering of circularly polarized photons off
longitudinally polarized electrons can be written~\cite{LT54} in the
laboratory frame as
 
\begin{equation}
\frac{d\sigma}{dE_\gamma} = \frac{d\sigma_0}{dE_\gamma}
\left[1 - P_\lambda P_e A_z(E_\gamma)\right],
\label{eq:sigma}
\end{equation}

where $d\sigma_0 / dE_\gamma$ is the unpolarized cross section, $E_\gamma$ is
the energy of the back-scattered Compton photons, $P_\lambda$ is the circular
polarization of the incident photons for the two helicity states $\lambda=\pm
1$, $P_e$ is the longitudinal polarization of the electron beam, and
$A_z(E_\gamma)$ is the longitudinal asymmetry function, which is shown in
Fig.~\ref{fig:analyzing-power.eps} for a 2.33~eV photon scattered off a
27.5~GeV electron. 

The total unpolarized cross section is 377~mb, and the differential cross
section is peaked at the maximum energy ($E_{\gamma,{\rm max}}=13.6$~GeV) of
the back-scattered Compton photons, hereafter called the Compton edge. The
longitudinal asymmetry function has a maximum of about 0.60 at the Compton
edge. As the energy of the Compton photons decreases, $A_z$ decreases rapidly
and becomes negative below 9.1~GeV, corresponding to scattering angles smaller
than $90^\circ$ in the electron rest frame, and returns to zero at $E_\gamma =
0$. Due to the enormous kinematic boost from the electron beam (the Lorentz
factor is $E_e/m_e \approx 5.4 \cdot 10^4$), most back-scattered Compton
photons are contained in a narrow cone centered around the initial direction
of the struck electrons. This is very advantageous to a polarization monitor
because it allows the detector to be far away (many tens of meters) from the
interaction region. However, the spatial distribution of the Compton photons
on the detector surface is given not only by the Compton kinematics but also
by the electron beam optics. Therefore, if the interaction point is chosen at
a position where the divergence of the electron beam is small, the transverse
size of the photon detector can be sufficiently small to accommodate only
moderate separations of the Compton photons and the electron beam, and thus
meet the spatial constraints given by the HERA electron ring.

\section{Apparatus} 

\begin{figure}[htbp]
\epsfig{figure=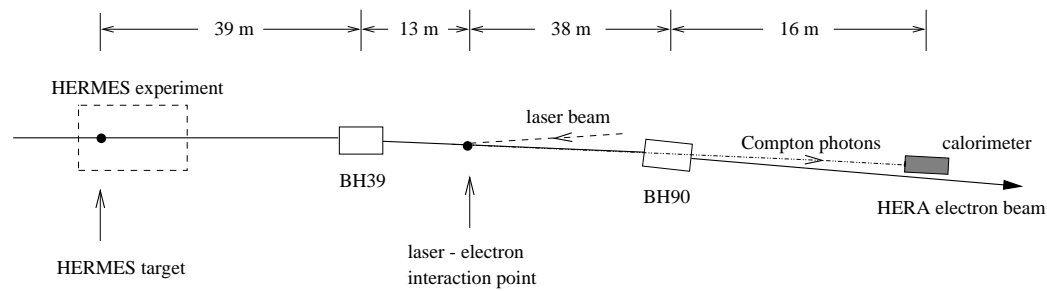,width=1.0\textwidth}
\vspace*{1mm}
\caption{Schematic overview of the Longitudinal Polarimeter in the HERA East section.}
  \label{fig:longpol_overview.eps}
\end{figure}

A schematic overview of the Longitudinal Polarimeter arrangement is shown in
Fig.~\ref{fig:longpol_overview.eps}. A circularly polarized photon beam from a
pulsed laser is focused on the HERA electron beam. The laser-electron
interaction point is located between the two bending magnets BH39 and BH90 at
39~m and 90~m from the HERMES target, respectively. A calorimeter measures the
energy of the back-scattered Compton photons for each laser pulse. If the
electron beam is longitudinally polarized, the energy distributions of the
Compton photons differ for left-handed and right-handed photon helicities.

\subsection{Laser and Optics}

\begin{figure}[htbp]
\epsfig{figure=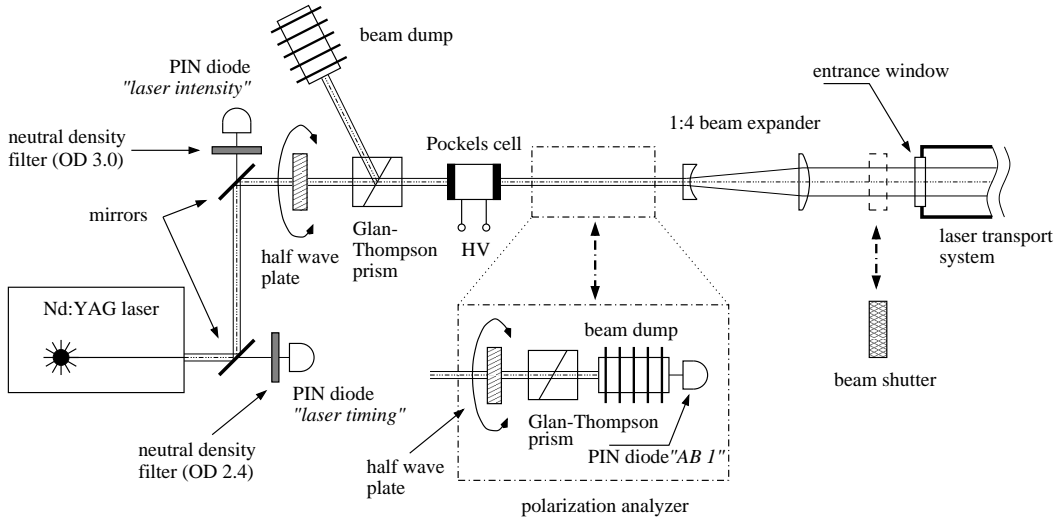,width=1.0\textwidth}
\vspace*{1mm}
\caption{Arrangement of the optical system in the laser room.}
  \label{fig:laser-layout.eps}
\end{figure}

A frequency-doubled, pulsed Nd:YAG laser~\cite{YAG} operated at 532~nm,
corresponding to a photon energy of $E_\lambda = 2.33$ eV, is used for the
measurements. The laser produces $3\, $ns long pulses of linearly polarized
light and can be operated with a continuously variable repetition rate from
single shot up to 100~Hz, and pulse energies from 1 to 250~mJ. The laser is
synchronized with the electron bunches in the HERA ring, and triggered at
close to 100~Hz. The timing and the intensity of each laser pulse are measured
by two photo diodes as shown in Fig.~\ref{fig:laser-layout.eps}. To minimize
pulse-to-pulse intensity fluctuations, the laser is operated at a fixed energy
of 100~mJ per pulse. The intensity of the laser pulses can be controlled by
passing the laser beam through a rotatable half wave plate and a fixed
Glan-Thompson prism. 

The linearly polarized laser light is converted to a circularly polarized beam
by passing it through an electrically reversible birefringent cell, known as a
Pockels cell. The voltage on the Pockels cell~\cite{Pockels} is adjusted to
produce a quarter wave phase shift which is reversed each pulse. The degree of
circular polarization $|P_\lambda|$ of the laser beam is larger than 0.999 for
each of the two helicity states $\lambda=\pm 1$, and is checked regularly with
a polarization analyzer~\cite{FB96} consisting of a rotatable half wave plate,
a Glan-Thompson prism, and a photo diode. Before entering the laser transport
system, the beam diameter is expanded by a factor of four by a set of
plano-concave and plano-convex lenses~\cite{CVI-beam-expander}. This beam
expander reduces the divergence of the laser beam to allow it to traverse the
$80\; $m long light path to the laser-electron interaction region and also to
reduce the resulting waist at the interaction point. In addition, it minimizes
the sensitivity to variations in the laser beam divergence, and it reduces the
energy density of the laser beam to prevent damage to the optical components
in the light path.

\begin{figure}[htbp]
\epsfig{figure=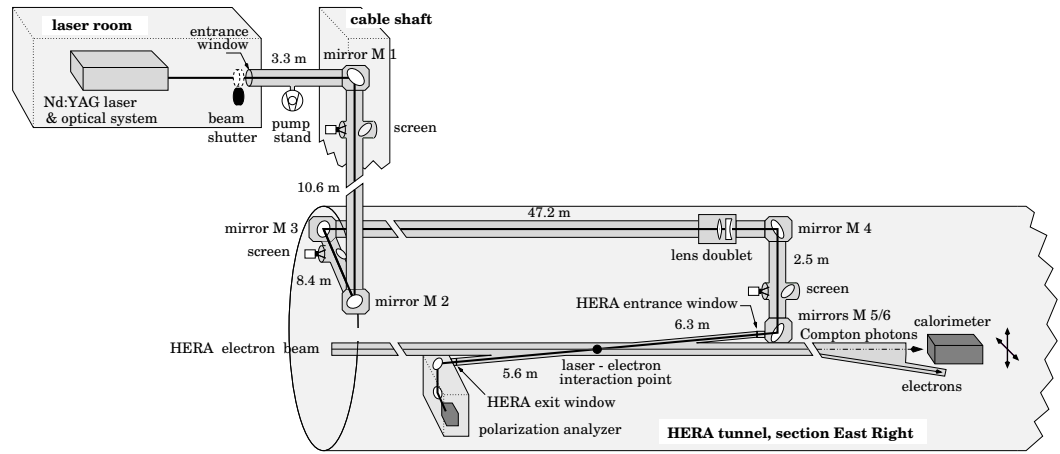,width=1.0\textwidth}
\vspace*{1mm}
\caption{Layout of the Longitudinal Polarimeter in the HERA East section.}
  \label{fig:lpol-layout.eps}
\end{figure}

The laser beam is guided by six remotely controlled mirrors~\cite{CVI-mirrors}
through a total of $72\, $m of stainless steel vacuum pipe, and focused with a
lens doublet~\cite{CVI-lens-doublett} on the HERA electron beam, as shown in 
Fig.~\ref{fig:lpol-layout.eps}. The mirrors are arranged in three
phase-compensated pairs to maintain the polarization of the photon beam close
to 100\,\%. Behind each mirror, a video camera is installed to monitor the
laser beam position. The laser beam enters the storage ring vacuum through a
$1\, $cm thick fused silica window~\cite{CVI-beam} and is brought into
collision with the electron beam at a vertical angle of 8.7~mrad. The window
was mounted with Helicoflex gaskets~\cite{helicoflex} to minimize stress such
that it has negligible optical retardation. 

At the interaction point, the laser spot has a diameter of approximately
0.5~mm, and the transverse size of the electron beam is $\sigma_x \approx $
0.6~mm horizontally and $\sigma_y \approx $ 0.2~mm vertically. Each electron
bunch is approximately $11\, $mm long (corresponding to 37$\, $ps), i.e. about
one hundred times shorter than the laser pulse. After passing through the
interaction point, the laser beam exits the storage ring vacuum system through
an identical vacuum window and enters a second polarization analyzer which
also monitors the position and the intensity of the laser light.

\subsection{Laser$-$Electron Interaction Region}

The location of the laser-electron interaction region was chosen to optimize
the rate of the back-scattered Compton photons versus the background rate, and
to minimize changes to the electron ring vacuum system. Maximizing the Compton
rate means that the crossing angle between the laser beam and the electron
beam should be as small as possible, and the horizontal widths of the electron
and laser beams should both be small. In addition, the transverse spatial
distribution of the back-scattered Compton photons due to the size and
divergence of the electron beam had to be minimized, since the back-scattered
photons have to travel about 54~m to the calorimeter.

The laser-electron interaction point is located in the East Right HERA tunnel
section, 13~m downstream of the first dipole magnet BH39, which bends the beam
by 0.54~mrad (Fig.~\ref{fig:longpol_overview.eps}). This is enough to prevent
a large fraction of the bremsstrahlung generated by the residual gas in the
long straight vacuum section upstream of BH39, and by the HERMES gas target in
particular, from reaching the calorimeter. On the other hand, it is little
enough that it rotates the spin by only 1.9$^\circ$. The corresponding
reduction of the measured longitudinal beam polarization is negligible
(0.06\,\%). 

The scattered electrons and photons travel with the unscattered electron beam
until the electrons are deflected by the dipole magnet BH90, which has a
bending radius of 1262~m and deflects the beam by 2.7~mrad. A collimator is
installed in the beam line 6~m downstream of BH90 to further reduce possible
bremsstrahlung contributions from the HERMES target. In order to minimize
changes to the electron ring vacuum system, the calorimeter position was
chosen 16~m downstream of BH90. This puts strict constraints on the transverse
size of the calorimeter, since the electron beam and the center of the
back-scattered Compton photon distribution are separated by only 42~mm at the
chosen position.

\subsection{The Compton Photon Detector} 

\begin{figure}[htbp]
\centerline{\epsfig{figure=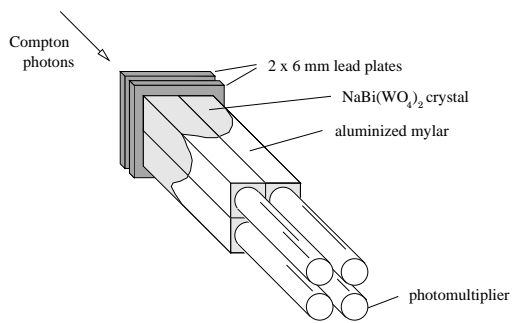,width=0.5\textwidth}}
\vspace*{1mm}
\caption{Schematic layout of the NaBi(WO$_4$)$_2$  crystal calorimeter.}
  \label{fig:crycalo-layout.eps}
\end{figure}

The detector assembly is mounted  on a remotely controlled table that can be
moved vertically and horizontally. A light tight aluminum box contains the
electromagnetic shower detector shown schematically in
Fig.~\ref{fig:crycalo-layout.eps}. The detector is positioned very close to
the electron beam pipe during normal operation. Therefore, the lateral face of
the box near the beam pipe is made of a $3\, $mm thick tungsten plate to
protect the detector against soft synchrotron radiation emerging from the beam
pipe.

The front of the detector is positioned 21 mm downstream of a copper vacuum
window in the HERA beam tube. This window is 2$\, $mm thick and 34$\, $mm in
diameter. The Compton photons enter the detector through a set of two $6\, $mm
($2 \times 1.1$ radiation lengths) thick lead plates, which serves as an
effective shield against the intense synchrotron radiation generated by the
dipole magnet BH90. The electromagnetic calorimeter consists of four optically
isolated NaBi(WO$_4$)$_2$ crystals. Each crystal is 20$\, $cm long (19
radiation lengths), 22$\, $mm wide and 22$\, $mm high, arranged in a $2\times
2$ array, as displayed in Fig.~\ref{fig:crycalo-layout.eps}. The crystal
material has a high index of refraction ($n=2.15$), and is very radiation-hard
($7\! \cdot \! 10^7$ rad)~\cite{NBW1,NBW2} and compact (Moli\`{e}re radius
2.38~cm). The Compton photons generate an electromagnetic shower in the lead
preshower and the crystals. The charged particles of the electromagnetic
shower produce \v{C}erenkov light, which is detected  by one photomultiplier
tube~\cite{PMT} for each crystal. The sharing of the shower between the four
calorimeter blocks allows a sub-millimeter alignment of the NaBi(WO$_4$)$_2$ 
array with respect to the Compton photon beam. 

\subsection{Trigger and Electronics}

The event trigger is provided by a pulser at a rate of approximately 200~Hz.
The laser is fired by only every second pulse, allowing a background event to
be recorded following each Compton event. Each laser pulse is synchronized
with the HERA bunch clock, which is provided by a bunch trigger module
(BTM)~\cite{BTM}. The BTM is also used to select a specific electron bunch in
a sequence determined by a programmable Digital Signal Processor
(DSP)~\cite{DSP}. Four consecutive events are recorded for each selected
bunch: for each of the two light helicity states, one background event and one
event where the laser was fired. The program of the DSP further provides the
option of scanning any subset of the beam bunches in any sequence. From the
recorded single bunch data, one can extract the polarization of a single
bunch, the average beam polarization of all the bunches, or the polarization
of any set of bunches, e.g. only colliding or non-colliding bunches. The
trigger also allows for pedestal or gain monitoring events during empty HERA
beam bunches. The HERMES gain monitoring system~\cite{HERMES-spectrometer}
monitors the response of the Compton photon detector by sending laser light
pulses through glass fibers that are coupled to the front faces of the
NaBi(WO$_4$)$_2$ crystals.

The signals from the four photomultiplier tubes are digitized by a charge
sensitive ADC~\cite{ADC}, and transferred by the DSP to the HERMES data
acquisition system, which is described in detail
elsewhere~\cite{HERMES-spectrometer}. Also the signals from the various
photodiodes are recorded. For each Compton event, the timing of the laser
pulse measured by the ``laser timing'' PIN diode (see
Fig.~\ref{fig:laser-layout.eps}) is recorded relative to the bunch timing by a
TDC~\cite{TDC}. 

\section{Polarimeter Operation}

Normal operation of the Longitudinal Polarimeter requires an optimum overlap
of the laser and electron beams in both space and time to maximize the
back-scattered Compton rate. The spatial overlap is achieved by steering the
laser beam horizontally through the interaction point with mirror M4 (see
Fig.~\ref{fig:lpol-layout.eps}). The timing of the laser pulse is set with
respect to the electron bunches by adjusting the laser trigger delay for
maximum luminosity. 

The luminosity is monitored continuously by the polarimeter control
system~\cite{SB96}. If the luminosity drops below a specified value, the
procedures described above are executed automatically to re-optimize it. This
online feedback system also ensures that the calorimeter remains centered on
the back-scattered Compton photon distribution, and that the startup and
shutdown of the laser system are executed automatically. Therefore, under
normal conditions, polarization measurements are performed without
intervention during HERA operation.

The detector can be operated in two different modes, the single-photon and the
multi-photon mode. In contrast to the single-photon mode, in which the energy
of each individual Compton photon is analyzed, in the multi-photon mode one
measures the total energy deposited in the detector by many Compton photons
per laser pulse interaction with an electron bunch. The multi-photon mode was
chosen as the standard mode of operation to provide high statistics single
bunch measurements in real time, and to overwhelm the bremsstrahlung
backgrounds originating from the residual vacuum pressure in the straight
section between the two dipole magnets BH39 and BH90. The single-photon mode
is used for test and diagnosis purposes only. 

\subsection{Single-Photon Mode}

\begin{figure}[htbp]
\epsfig{figure=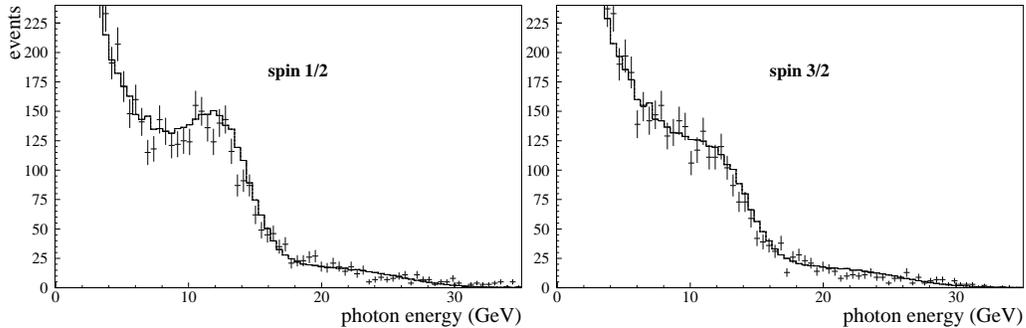,width=1.0\textwidth} 
\vspace*{1mm}
\caption{Energy spectra collected in single-photon mode for the
spin-$\frac{1}{2}$ and spin-$\frac{3}{2}$ configurations at a beam
polarization of 0.51. The solid line is the result of a
simulation~\cite{Zomer} for a Compton  (bremsstrahlung) rate of 0.02 (0.06)
per bunch.} 
\label{fig:single-photon.eps} 
\end{figure}

The advantages of running in single-photon mode would be twofold. The
asymmetries are large, up to 0.60 at the Compton edge (see
Fig.~\ref{fig:analyzing-power.eps}), and  the energy spectra can be compared
to the Compton cross sections. Operation of the Longitudinal Polarimeter in
this mode is possible if the laser pulse intensity is drastically reduced.
However, the resolution for single-photon events is rather poor, as shown in
Fig.~\ref{fig:single-photon.eps}, because most of the generated \v{C}erenkov
light is trapped in the crystals and does not reach the photomultiplier tubes
due to the high index of refraction of the crystals and the $3\, $mm air gap
between the crystals and the photomultiplier tubes. While Compton spectra can
be produced, and the beam polarization can be extracted, this is not a
feasible mode of operation with the 100~Hz laser, since a measurement of the
beam polarization with an absolute statistical accuracy of 0.01 takes about
2.5 hours. In comparison, such a measurement takes only one minute in the
multi-photon mode.

In the single-photon mode, the asymmetry can be written as 

\begin{equation}
A_s (E_\gamma) = \frac{(d\sigma/dE_\gamma )_{\frac{1}{2}} -
(d\sigma/dE_\gamma )_{\frac{3}{2}}}{(d\sigma/
dE_\gamma )_{\frac{1}{2}} + (d\sigma/dE_\gamma )_{\frac{3}{2}}} 
=  P_c P_e A_z(E_\gamma),
\label{eq:asymmetry}
\end{equation}

where  $({d\sigma/dE_\gamma})_{\frac{1}{2}}$ and $({d\sigma/{\rm
d}E_\gamma})_{\frac{3}{2}}$ are the cross sections for the electron-photon
configurations where the incident spins are antiparallel and parallel,
respectively, and $P_c=\frac{1}{2}\; (|P_{+1}|+|P_{-1}|)$ is the average
circular light polarization.

The electron beam polarization is determined by fitting the energy spectra for
the two spin configurations using a simulation (solid line in
Fig.~\ref{fig:single-photon.eps}) that includes the response function and
resolution of the detector, and realistic background conditions~\cite{Zomer}
above a Compton photon energy of 4~GeV. Whereas the simulation represents the
data well above 4~GeV, it considerably underestimates the background at lower
energies.

\subsection{Multi-Photon Mode}

The operation of the Longitudinal Polarimeter in multi-photon mode has the
advantage of being effectively independent of bremsstrahlung background in the
HERA storage ring. A large number of Compton photons is produced each time a
laser pulse interacts with an electron bunch. These photons are detected
together by the calorimeter, which measures their energy sums
$I_{\frac{1}{2}}$ and $I_{\frac{3}{2}}$ for the  \mbox{spin-$\frac{1}{2}$} and
\mbox{spin-$\frac{3}{2}$} electron-photon configurations, respectively. In the
multi-photon mode, an energy asymmetry is formed as

\begin{equation}
A_m  =  \frac{I_{\frac{1}{2}}-I_{\frac{3}{2}}}{I_{\frac{1}{2}}+I_{\frac{3}{2}}} =  
P_c P_e A_p ,
\label{eq A_e_av}
\end{equation} 

where $A_p$ is the analyzing power of the process. Under the assumption that
the photomultiplier signals are linear over the full single-photon to
multi-photon operating range, $A_p$ is given by the integrals over the energy
weighted cross sections for the \mbox{spin-$\frac{1}{2}$} and
\mbox{spin-$\frac{3}{2}$} configurations and $P_c P_e=1$, multiplied by the
single-photon relative response function $r(E_\gamma) = S(E_\gamma)/E_\gamma$ 
(where $S$ is the digitized ADC signal) of the detector. The analyzing can be
written as

\begin{eqnarray}
A_p  &=& 
\frac{\Sigma_{\frac{1}{2}}-\Sigma_{\frac{3}{2}}}{\Sigma_{\frac{1}{2}}+
\Sigma_{\frac{3}{2}}} , \\ \nonumber
\label{eq A_p}
{\rm with}\qquad   \Sigma_{i} &=& \int_{E_{\gamma,{\rm min}}}^{E_{\gamma,{\rm
max}}} \!\! (d\sigma/dE_\gamma )_{i}\;  E_\gamma \; r(E_\gamma) \; dE_\gamma,
\qquad   \mbox{\small $i=\frac{1}{2} \, , \; \frac{3}{2}$}.
\end{eqnarray} 

\begin{figure}[htbp] 
\centerline{\epsfig{figure=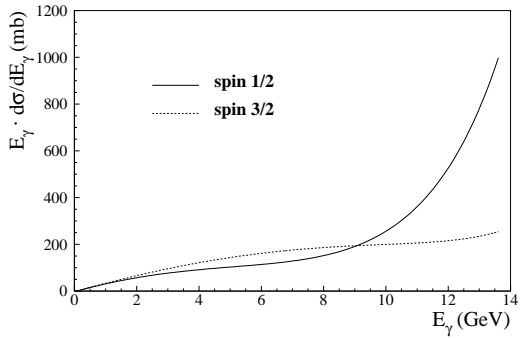,width=0.5\textwidth}}
\vspace*{1mm} 
\caption{Energy-weighted cross sections for the \mbox{spin-$\frac{1}{2}$}
(solid curve) and \mbox{spin-$\frac{3}{2}$} (dashed curve) configurations.} 
\label{fig:esec.eps}   
\end{figure}

Assuming a linear energy response of the detector, the analyzing power has a
value of 0.1838 for $E_\lambda = 2.33 \,$eV and $E_e = 27.5\,$GeV. The
energy-weighted Compton cross sections for the two spin configurations are
shown in Fig.~\ref{fig:esec.eps}. Their asymmetry is largest for photon
energies close to the Compton edge ($E_{\gamma,{\rm max}}$). For small photon
energies the two distributions are nearly identical. This has the advantage
that the analyzing power is not very sensitive to the detector energy
threshold $E_{\gamma,{\rm min}}$. 

\subsection{Polarization Determination}
\label{sec:pol-det}

\begin{figure}[htbp]
\centerline{\epsfig{figure=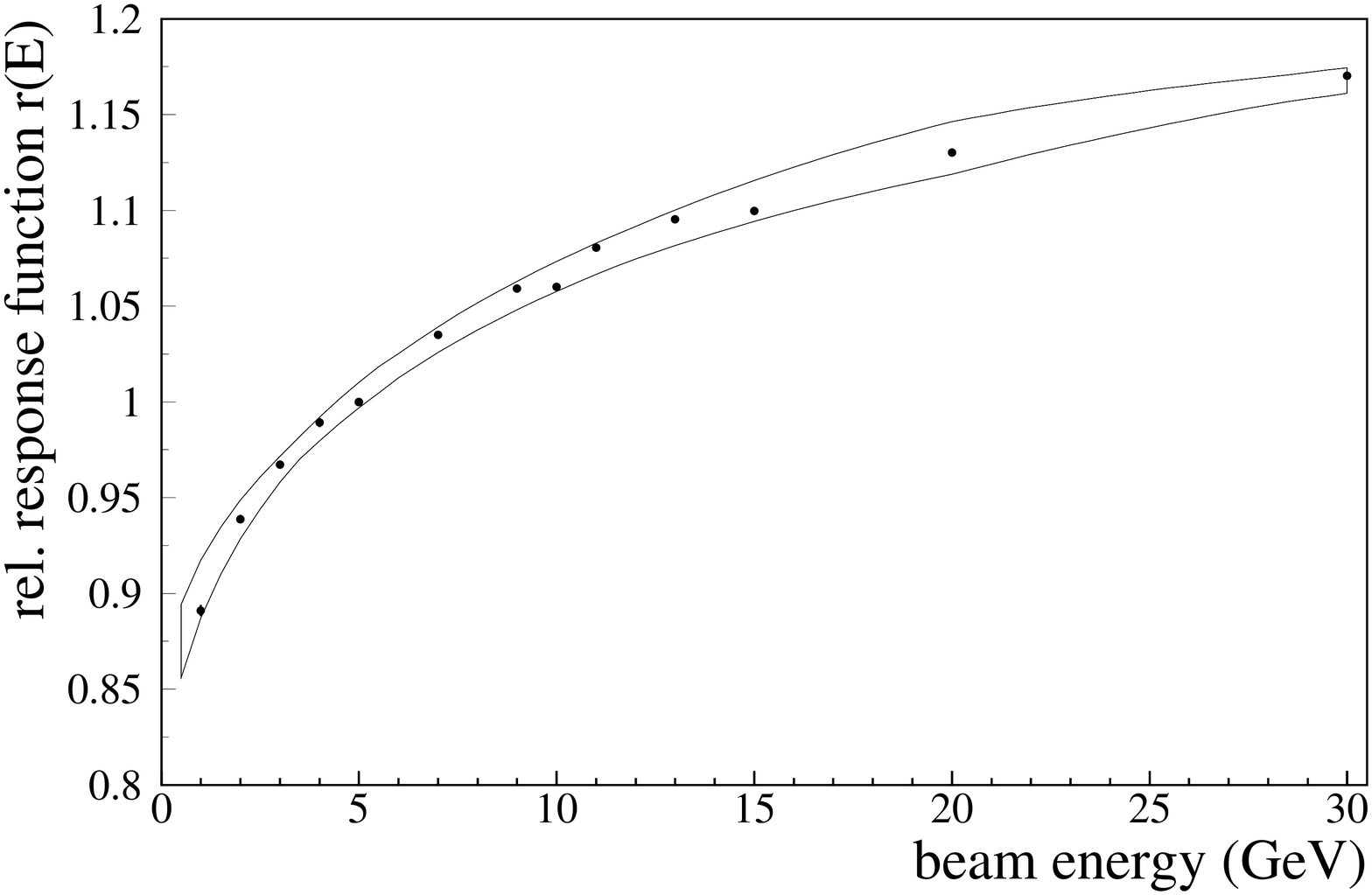,width=0.5\textwidth}}
\vspace*{1mm}
\caption{Relative calorimeter response function, normalized to unity at 5 GeV,
as determined in the DESY T22 and CERN X5 test beams. The band represents the
systematic uncertainty.} 
\label{fig:response-function.eps}  
\end{figure}

In order to determine the single-photon relative response function
$r(E_\gamma)$ of the ${\rm NaBi(WO_4)_2}$ calorimeter, test beam measurements
were performed at DESY and CERN, covering the entire energy range of the
back-scattered Compton photons, as shown in
Fig.~\ref{fig:response-function.eps}. The resulting analyzing power was found
to be $0.1929 \pm 0.0017$. Note that only the precision of the measurement of
the calorimeter response is important, and not its deviation from linearity.

\begin{figure}[htbp]
\centerline{\epsfig{figure=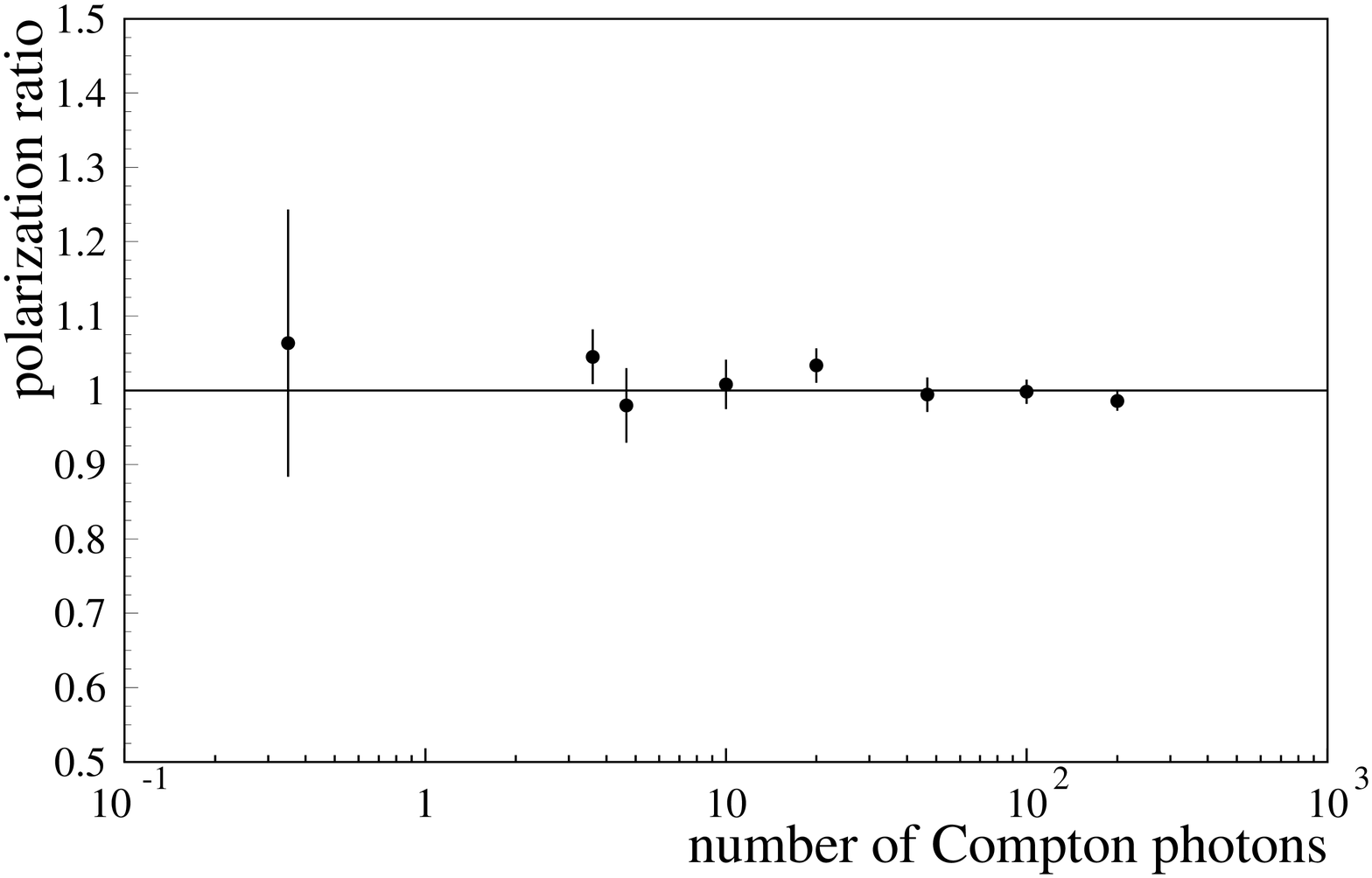,width=0.5\textwidth}}
\vspace*{1mm}
\caption{Ratio of longitudinal to transverse electron beam polarization as a function of
the average number of Compton photons detected in the calorimeter.} 
\label{fig:single-to-multi-mode.eps}  
\end{figure}

However, to apply the result for the relative response function obtained from
the test beam measurements to Eq.~(6), it was necessary to show that the
photomultiplier response is linear over the full single-photon to multi-photon
operating range. This was verified using the gain monitoring system. The
ultimate test was then to show that the measurement of the beam polarization
was not affected by changing from single-photon to multi-photon mode. This was
demonstrated by attenuating the laser beam intensity over three orders of
magnitude using a rotatable half wave plate and a fixed Glan-Thompson prism
while increasing the photomultiplier high-voltage such that the digitized
signal of the photomultiplier remained constant. The test was performed during
stable electron beam conditions while monitoring the beam polarization with
the Transverse Polarimeter by observing the ratio of the polarization of the
two polarimeters. As shown in Fig.~\ref{fig:single-to-multi-mode.eps}, the
ratio was constant over the entire range. 

\begin{figure}[htbp]
\centerline{\epsfig{figure=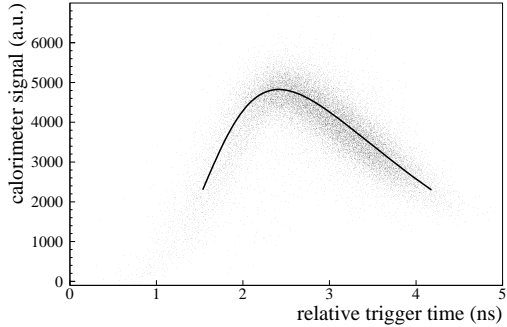,width=0.5\textwidth}}
\vspace*{1mm}
\caption{Temporal profile of the laser pulses as sampled by an electron bunch.
The solid line through the distribution is a fit which is used to correct the
calorimeter response.} 
\label{fig:lumi-corr.eps}  
\end{figure}

Once the calorimeter response was understood in the single-photon and
multi-photon modes, the longitudinal beam polarization was determined by
evaluating the calorimeter signals for every bunch individually.  Although the
laser is triggered by a precise electronic signal that is synchronized with
the HERA bunch timing, the time of the resulting light pulse fluctuates within
$\pm 1.5\, $ns. Because of this fluctuation and the finite crossing angle, the
$37\, $ps long electron bunches interact with varying parts of the $3\, $ns
long laser pulses. As mentioned earlier, the timing of each laser pulse is
recorded relative to the trigger signal. The calorimeter signal reflects the
temporal profile of the laser pulses if it is plotted versus the relative
trigger time, as shown in Fig.~\ref{fig:lumi-corr.eps}. With a fit to this
distribution, the calorimeter response is corrected for this variation.

\begin{figure}[htbp]
\centerline{\epsfig{figure=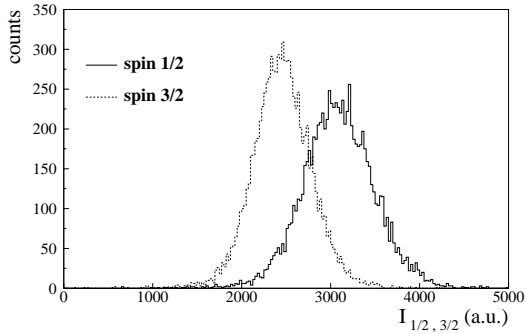,width=0.5\textwidth}}
\vspace*{1mm}
\caption{Spectra collected in multi-photon mode for the
\mbox{spin-$\frac{1}{2}$} (solid histogram) and \mbox{spin-$\frac{3}{2}$}
(dashed histogram) configurations for a specific electron bunch with a beam
polarization of 0.59.} 
\label{fig:multi-photon.eps}  
\end{figure}

Switching between the two light helicity states results in the two energy
distributions for the corrected calorimeter signals $I_{\frac{1}{2}}$ and
$I_{\frac{3}{2}}$, displayed in Fig.~\ref{fig:multi-photon.eps} for an
individual bunch. The longitudinal polarization of each electron bunch is
determined from the asymmetry of the means of these two energy distributions
divided by the analyzing power and the measured circular light polarization
(Eq.~(\ref{eq A_e_av})). This calculation is provided every minute. The
longitudinal beam polarization is finally computed as the mean of the
individual bunch polarizations weighted by the corresponding time-averaged
bunch currents.

\subsection{\v{C}erenkov Light Attenuation}
\label{sec:light-att}

A large number of Compton photons can be produced per laser pulse when the
polarimeter is operated in the multi-photon mode, ranging from a few photons
to many thousand. During normal operating mode in which about 1000
back-scattered Compton photons are produced at the beginning of a fill,
approximately 250 times more energy is deposited in the calorimeter than the
highest energies (bremsstrahlung) deposited in the single-photon mode, since
the average energy deposited per Compton photon is 6.8 GeV. In order to
attenuate the \v{C}erenkov light to protect the photomultiplier tubes from
saturation, a remotely controlled movable, perforated nickel foil could be
inserted into the 3~mm air gap between the NaBi(WO$_4$)$_2$ crystals and the
photomultiplier tubes. This was initially the standard mode of operation.

Even though the NaBi(WO$_4$)$_2$ crystals are 19~radiation lengths long, there
is a small amount of longitudinal shower leakage into the photomultiplier
tubes. Unfortunately, the corresponding shower particles generate a large
signal in the photomultiplier tubes, which introduces a substantial
non-linearity in the energy response. The longitudinal shower leakage signal
derives mostly from the highest energy Compton photons and hence has a large
analyzing power. When the \v{C}erenkov light produced in the NaBi(WO$_4$)$_2$
crystals was attenuated by the nickel foil, the shower leakage signal
dominated the signal in the photomultiplier tubes. This altered the response
function and increased the analyzing power of the detector by about 25\,\%.

The polarimeter has been operated without any light attenuators since early
1999. This was also the case for the test beam calibrations. The gain in the
photomultiplier tubes has to be reduced in the multi-photon mode by about a
factor of 200. As described in Section~\ref{sec:pol-det}, it was verified that
the photomultiplier tubes are linear over this large range in gain. To address
concerns about long-term stability, linearity, and radiation damage, a
tungsten/scintillator sampling calorimeter, similar to the one employed in the
Transverse Polarimeter~\cite{Bar93} but without position sensitivity, is moved
in the Compton photon beam periodically. It acts as an independent device to
check the beam polarization measurement and is otherwise not exposed to
bremsstrahlung and direct synchrotron radiation.

\section{Polarimeter Performance}
\label{sec:Results}

\begin{figure}[htbp]
\centerline{\epsfig{figure=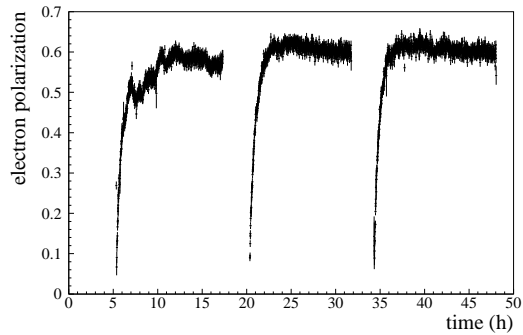,width=0.5\textwidth}}
\vspace*{1mm}
\caption{Longitudinal beam polarization versus time for three consecutive
fills.}
\label{fig:lpol.eps}  
\end{figure}

Since early 1997 the Longitudinal Polarimeter has routinely measured the HERA
electron beam polarization for the HERMES experiment. Typical electron beam
fills last from eight to twelve hours, starting with a ramped injection
current of 40 to 45~mA and ending usually with a controlled beam dump when the
current reaches about 10~mA. Fig.~\ref{fig:lpol.eps} shows an example of
polarization measurements as a function of time for three consecutive fills.
Each data point represents a one minute measurement of the longitudinal
polarization with an absolute statistical accuracy of 0.01. The time structure
in the first fill displayed in Fig.~\ref{fig:lpol.eps} is the result of tuning
efforts by the HERA operators.  In the 1997 and 1998 running periods the
electron beam helicity was reversed every few months; since then it has been
reversed once a month. 

\subsection{Single Bunch Measurements}

\begin{figure}[htbp]
\centerline{\epsfig{figure=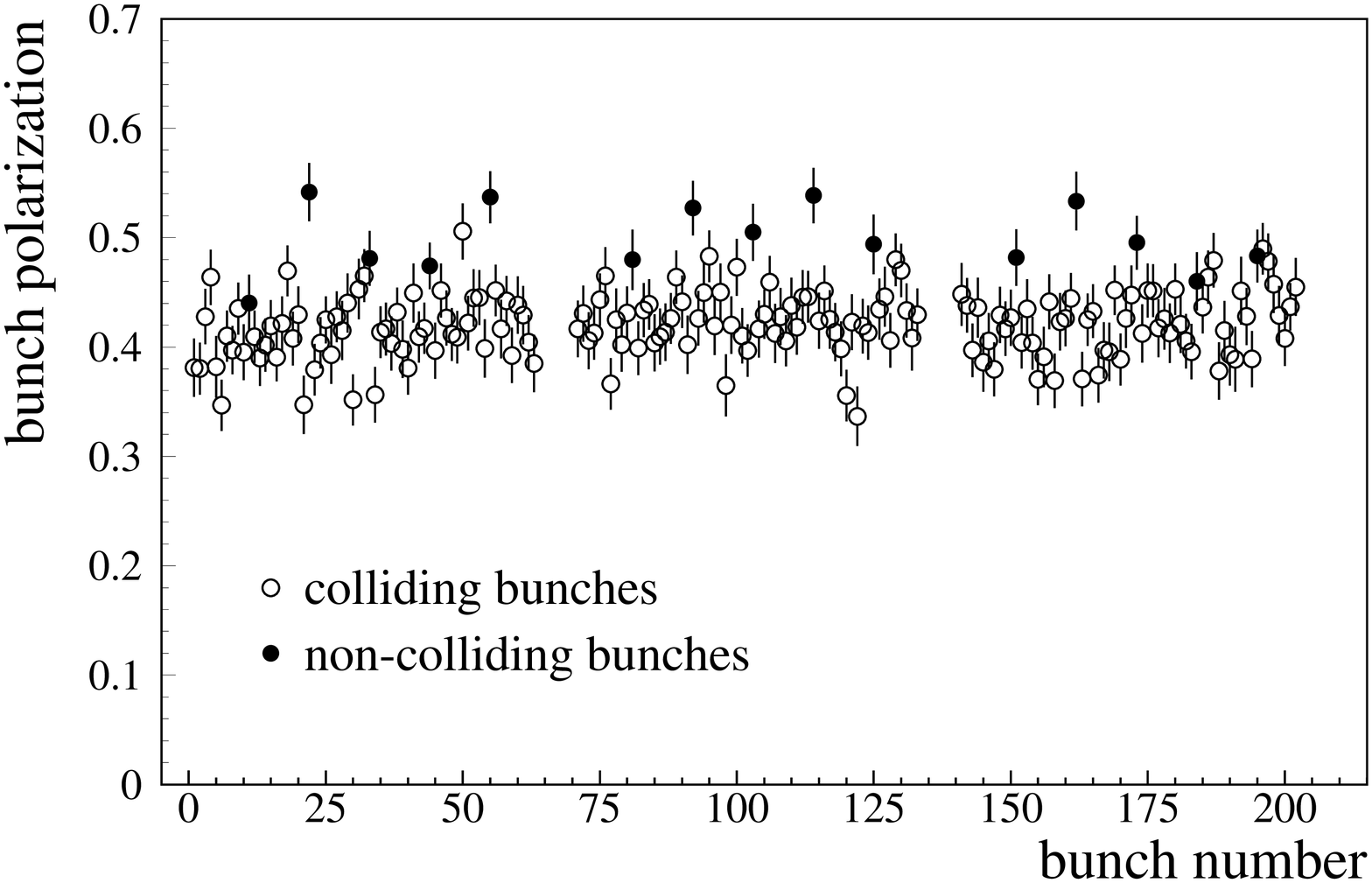,width=0.5\textwidth}}
\vspace*{1mm}
\caption{Polarization of the individual beam bunches, as measured by the
Longitudinal Polarimeter.}
\label{fig:sbunch.eps}  
\end{figure}

\begin{figure}[htbp]
\centerline{\epsfig{figure=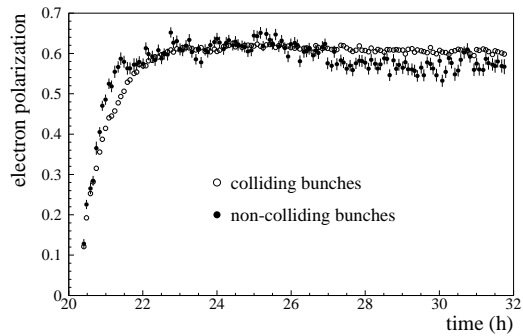,width=0.5\textwidth}}
\vspace*{1mm}
\caption{Longitudinal polarization of the colliding and non-colliding beam
bunches versus time, for the second beam fill in  Fig.~\ref{fig:lpol.eps}.}
\label{fig:coll-noncoll.eps}  
\end{figure}

The Longitudinal Polarimeter measures the polarization of individual bunches,
as shown in Fig.~\ref{fig:sbunch.eps}. Each data point represents a
measurement lasting twenty minutes with an absolute statistical accuracy of
0.03. Not all electron bunches collide with proton bunches in HERA, and it was
found that the colliding and non-colliding electron bunches can have different
polarization values. This is believed to be caused by beam-beam interactions
between the electron and proton beams and the associated tune shifts.
Comparison of the polarization of the 174 colliding and the 15 non-colliding
bunches is a useful tool for tuning the accelerator to optimize polarization.
This information is shown in Fig.~\ref{fig:coll-noncoll.eps} and is provided
in real time to the HERA control room every minute, with an absolute
statistical precision of 0.01 (0.04) for the colliding (non-colliding)
bunches. 

Analyzing individual electron bunches is as of yet unique to the Longitudinal
Polarimeter. An upgrade of the data acquisition system~\cite{Pol-2000} of the
Transverse Polarimeter at HERA, which is in progress, will also have this
important feature.  This detailed polarization information about the electron
beam will be crucial for the collider experiments H1 and ZEUS since they are
preparing to measure spin observables after the luminosity upgrade in 2001.
Whereas the HERMES experiment is sensitive to the average beam polarization of
all the bunches, the collider experiments are sensitive to the colliding
bunches only. 

\subsection{Systematic Uncertainties}

Various studies have been performed to investigate the systematic uncertainty
associated with the polarization measurements by the Longitudinal Polarimeter
operated in the multi-photon mode. Since the polarization of the electron beam
is obtained from a measurement of an asymmetry $A_m$ (see Eq.~(\ref{eq
A_e_av})), potential sources of false asymmetries were investigated. Other
studies quantified the precision with which the analyzing power $A_p$ and the
circular light polarization $P_c$ at the interaction point are determined. The
various contributions to the total systematic uncertainty are summarized in
Table 1, and apply to the polarimeter operating conditions without optical
filters in the calorimeter, i.e. since early 1999.

The largest contribution to the overall systematic uncertainty originates from
the determination of the analyzing power, which depends strongly on the exact
shape of the relative detector response function $r(E_{\gamma})$ (see
Fig.~\ref{fig:response-function.eps}). The energy calibration of the detector
in the test beams takes into account most sources that can lead to a
non-linear response of the detector including the crystals, and the signal
generated in the photomultiplier tubes by the longitudinal shower leakage (see
section~\ref{sec:light-att}). It does not account for the response of the
photomultiplier tubes in the multi-photon mode, but it accounts for the
low-energy cut-off from the lead absorber, and the limited size of the
calorimeter. The analyzing power was determined to a precision of $1.2\,\%$
that was calculated by propagating the systematic uncertainty of the relative
response function ($0.9\,\%$) shown in Fig.~\ref{fig:response-function.eps},
and by including the uncertainty of the  transition from single-photon to
multi-photon mode ($0.8\,\%$) shown in
Fig.~\ref{fig:single-to-multi-mode.eps}.

The long-term stability of the detector response function is checked by
monitoring the sources that can produce a time-dependent non-linear detector
response. The linearity of the photomultiplier tubes is checked continuously
with the gain monitoring system over the full multi-photon operating range and
is found to deviate by less than $0.4\,\%$ over the annual running period. The
effect on the analyzing power is of the same size. The annual radiation dose
deposited in the crystals was determined to be about ten times below the level
of damage. Instead of considering all contributions separately, the overall
systematic uncertainty can also be estimated by periodically performing
measurements with the sampling calorimeter to compare the polarization
measurements of the two detectors. Based on this comparison, the systematic
uncertainty associated with the long-term instability of the analyzing power
is $0.5\,\%$.

Possible false asymmetries introduced by gain-mismatched photomultiplier tubes
have been considered. An iterative method based on data from scanning the
Compton photon beam across the detector front face is used to gain-match the
detector elements within an accuracy of about 5\%. The gains of the
photomultiplier tubes are monitored continuously and found to change by
approximately $20-30\,\%$ during a beam fill, returning to their initial
values between fills. These short-term drifts differ by only a few percent for
the four photomultiplier tubes and therefore have no net effect on the
polarization measurement, since the Longitudinal Polarimeter does not depend
on an absolute energy calibration. Long-term deviations of the relative gains
during the annual running were found to affect the beam polarization
measurement by less than $0.3\,\%$.

The circular polarization of the laser light can be determined very
precisely~\cite{MB00} immediately following the Pockels cell in the laser
room. However, it can in principle be different at the interaction point,
given the fact that the laser beam has to be passed through windows and lenses
and be reflected from mirrors before it interacts with the electron beam. To
estimate this uncertainty, the laser beam polarization was measured after the
storage ring vacuum window with the identical analyzer that is normally
mounted in the laser room. The two vacuum windows were removed, then each
window was mounted separately, and finally the ring vacuum was re-established,
while measuring the laser beam polarization after each step. Based on these
measurements, a systematic uncertainty of $0.2\,\%$ was assigned to the
circular polarization of the laser light at the interaction point.

The measurement may also be affected by changes in the phase space of the
laser beam at the interaction point due to an imperfectly  aligned Pockels
cell. Horizontal or vertical shifts of the laser beam can occur when the
voltage across the cell is changed, resulting in a helicity-dependent
luminosity and hence a false energy asymmetry. To quantify extraneous 
helicity-dependent beam shift effects in the system, we performed two tests.
First, a half wave plate was temporarily mounted immediately following the
Pockels cell. Except for the expected change of sign in the measurement of the
electron beam polarization, no change in the magnitude was observed within the
$0.3\,\%$ precision of the test. However, this test does not account for
non-optimal laser and electron beam overlap. This is important since the
sensitivity to a helicity-dependent laser beam shift increases with
decreasing  overlap. Therefore a second test was performed by changing the
overlap of the two beams within the limits of the normal operating conditions.
This showed that the impact on the energy asymmetry is at most $0.3\,\%$.
Combining the two values leads to a total contribution of $0.4\,\%$.

The position and size of the Compton photon beam incident on the calorimeter
is determined by the electron beam orbit conditions at the interaction point.
During normal HERA luminosity operation, variations of the  size and
divergence of the electron beam are so small that the impact on the
calorimeter response is negligible. However, a change of the position or slope
of the electron beam at the interaction point can result in  a shift of the
Compton photon distribution away from the center of the calorimeter. In these
cases, the online feedback system of the polarimeter automatically
repositions  the calorimeter center on the Compton photon beam to better than
1~mm precision. By scanning the Compton photon beam across the calorimeter
front face, it has been determined that within the relevant operating range,
the effect on the measurement of the beam polarization is less than $0.6\,\%$.
To estimate the effect of slow beam drifts during a fill, the slope of the
electron beam was moved over the maximal observed range while keeping the
Compton photon distribution centered on the calorimeter. No influence on the
polarization measurement was observed within the $0.5\,\%$ accuracy of the
study. Combining the uncertainties of the two tests leads to a total
contribution of at most $0.8\,\%$.

\begin{table}[htbp]
\caption{The various contributions to the fractional systematic uncertainty of
the longitudinal electron beam polarization $P_e$.}
\begin{center}
\begin{tabular}{lr}
\hline
Source of systematic uncertainty & $\Delta P_e\, / \, P_e$  \\
\hline
Analyzing power       			          & $\pm\,1.2\,\%$ \\
Analyzing power long-term instability             & $\pm\,0.5\,\%$ \\
Gain mismatching                                  & $\pm\,0.3\,\%$ \\
Laser light polarization                          & $\pm\,0.2\,\%$ \\
Pockels Cell misalignment                         & $\pm\,0.4\,\%$ \\
Electron beam instability        		  & $\pm\,0.8\,\%$ \\
				{\bf Total}       & {$\bf \pm\,1.6\,\%$}  \\
\hline
\end{tabular}
\end{center}
\noindent
\label{ta:syst-errors}
\end{table}

The various contributions to the systematic uncertainties of the Longitudinal
Polarimeter have been considered separately and added in quadrature to a total
uncertainty of $1.6\,\%$ (see Table~\ref{ta:syst-errors}). Those systematic
uncertainties of the two HERA electron beam polarimeters that relate to
stability and reproducibility (not absolute scale) can be further studied by
comparing their performances over an extended period of time. Non-statistical
fluctuations in the ratio of their results over the 1999-2000 running periods
correspond to a relative systematic stability of $\sigma = 1.6\,\%$, which is
compatible with the quadratic sum of contributions estimated from the two
instruments.

\section{Summary}

We have designed and constructed a Compton back-scattering laser polarimeter
which routinely measures the longitudinal polarization of the HERA electron
beam for the HERMES experiment. The Longitudinal Polarimeter determines the
beam polarization with an absolute statistical precision of 0.01 per minute
and a fractional systematic uncertainty of $1.6\,\%$. The polarimeter also
measures the polarization of individual electron bunches, a feature that is
currently not available to the Transverse Polarimeter. It was found that the
individual bunches can each have a significantly different polarization. This
observation can be further analyzed if one groups the bunches into colliding
and non-colliding bunches. The variation and the time evolution of the
polarization of the individual bunches and of classes of bunches provide
important additional information for achieving high beam polarization at HERA.

\section{Acknowledgments}

We would like to thank R.~Fastner and the crew of the machine shop from the
University of Freiburg for the design and construction of the optical
transport system, and G.~Braun for his help in designing some of the
electronic components. We thank M.~Spengos and E.~Steffens for their help in
the initial planning of the project, N.~Meyners for his design of the safety
interlock system, and H.D.~Bremer for his help in the design and construction
of the calorimeter and the calorimeter table. We also thank T.~Behnke,
A.~Miller, and  P.~Sch\"uler for many useful discussions and their comments to
this article. We are grateful to E.~Belz, O.~H\"ausser, R.~Henderson, M.~Ruh,
M.~Woods, and R.~Zurm\"uhle for their help and advice. We acknowledge the DESY
management for its support, and the DESY staff for the significant effort in
the planning, design, and construction of the Longitudinal Polarimeter. We
especially acknowledge the efforts of the HERA machine group to deliver high
beam polarization. This work was supported by the German Bundesministerium
f\"ur Bildung, Wissenschaft, Forschung und Technologie, and the US National
Science Foundation.


\begin{thebibliography}{99}
\bibitem{BTM} 
The bunch trigger module is developed by the MKI group at DESY.
\end{thebibliography}
\end{document}